\documentclass[%
 aip,apl,showplace,
 amsmath,amssymb,
 reprint,%
]{revtex4-1}
\usepackage{graphicx}% Include figure files
\usepackage{dcolumn}% Align table columns on decimal point
\usepackage{bm}% bold math
\begin{document}

\preprint{AIP/123-QED}

\title{Enhanced Directional Coupling of Light with a Whispering Gallery Microcavity}

\author{Fuchuan Lei}
\email{fuchuan.lei@oist.jp}
\affiliation{Light-Matter Interactions for Quantum Technologies Unit, Okinawa Institute of Science and Technology Graduate University, Onna, Okinawa 904-0495, Japan}%Lines break automatically or can be forced with \\
\author{Georgiy Tkachenko}%
\affiliation{Light-Matter Interactions for Quantum Technologies Unit, Okinawa Institute of Science and Technology Graduate University, Onna, Okinawa 904-0495, Japan}%
\author{Xuefeng Jiang}%
\affiliation{Department of Electrical and Systems Engineering, Washington University, St. Louis, Missouri 63130, USA}%
\author{Jonathan M. Ward}%
\affiliation{Light-Matter Interactions for Quantum Technologies Unit, Okinawa Institute of Science and Technology Graduate University, Onna, Okinawa 904-0495, Japan}%
\author{Lan Yang}
\affiliation{Department of Electrical and Systems Engineering, Washington University, St. Louis, Missouri 63130, USA}
\author{S\'ile Nic Chormaic}%
 \affiliation{Light-Matter Interactions for Quantum Technologies Unit, Okinawa Institute of Science and Technology Graduate University, Onna, Okinawa 904-0495, Japan}%
 \affiliation{Universit\'e Grenoble Alpes, CNRS, Grenoble INP, Institut N\'eel, 38000 Grenoble, France}
\email{sile.nicchormaic@oist.jp}

\date{\today}% It is always \today, today,

\begin{abstract}
Directional coupling of light in nanophotonic circuits has recently attracted increasing interest, with numerous experimental realizations based on broken rotational or mirror symmetries of the light-matter system. The most prominent underlying effect is the spin-orbit interaction of light in subwavelength structures. Unfortunately, coupling of light to such structures is, in general, very inefficient. In this work, we experimentally demonstrate an order of magnitude enhancement of the directional coupling between two nanowaveguides by means of a whispering gallery microcavity. We also show that both transverse magnetic and transverse electric modes can be used for the enhancement.
\end{abstract}

%\setboolean{displaycopyright}{true}

%\begin{document}

\maketitle
Engineering photon emission and scattering at subwavelength scales is central to many applications ranging from near-field microscopy to communication and quantum technologies. It is well known that emission and scattering are not solely determined by the light source, but also by its surrounding medium and electromagnetic environment~\cite{novotny2012principles}. For example, the lifetime of dipole emitters or the scattering cross-section of nanoparticles can be modified by micro- and nanocavities~\cite{noda2007spontaneous,kippenberg2009purcell,zhu2014interfacing,ward2019excitation}, while the radiation directionality can be controlled by nanoantennae~\cite{curto2010unidirectional,lee2011planar}. In contrast to macroscale and free-space optical systems, geometries with strongly confined fields (e.~g. nanowaveguides or on-chip photonic circuits) feature an interaction between the spin and orbital angular momenta of light~\cite{bliokh2015spin}. This effect prompted the development of various photonic devices based on directional coupling and channeling of chiral light using plasmonic~\cite{lee2012role,lin2013polarization,rodriguez2013near} or dielectric~\cite{luxmoore2013interfacing,junge2013strong,neugebauer2014polarization,rodriguez2014resolving} material structures,  in both the classical~\cite{petersen2014chiral,le2015nanophotonic,shao2018spin,PhysRevApplied.11.064041} and the quantum~\cite{mitsch2014quantum,shomroni2014all,sollner2015deterministic,coles2016chirality} regimes.

Despite the above achievements, the demonstrated directional optical couplers are very inefficient. For example, about 10\% of light \textit{scattered} by a dipole-like nanoparticle can be collected by an optical nanofiber, as estimated in~\cite{petersen2014chiral}.  However, if one considers the incident optical power, $P_{in}$, coupled to the fiber and the output signal, $P_{out}$, channeled through the fiber, the efficiency is $\eta=P_{out}/P_{in}\sim10^{-8}$ (calculated from the photon counts and the experimental parameters given in the supplementary material for~\cite{petersen2014chiral}). Because the efficiency is so small, output signals from nanophotonic couplers are typically measured via single-photon  detection.

\begin{figure}[t]
\centering
\includegraphics[width=0.9\linewidth]{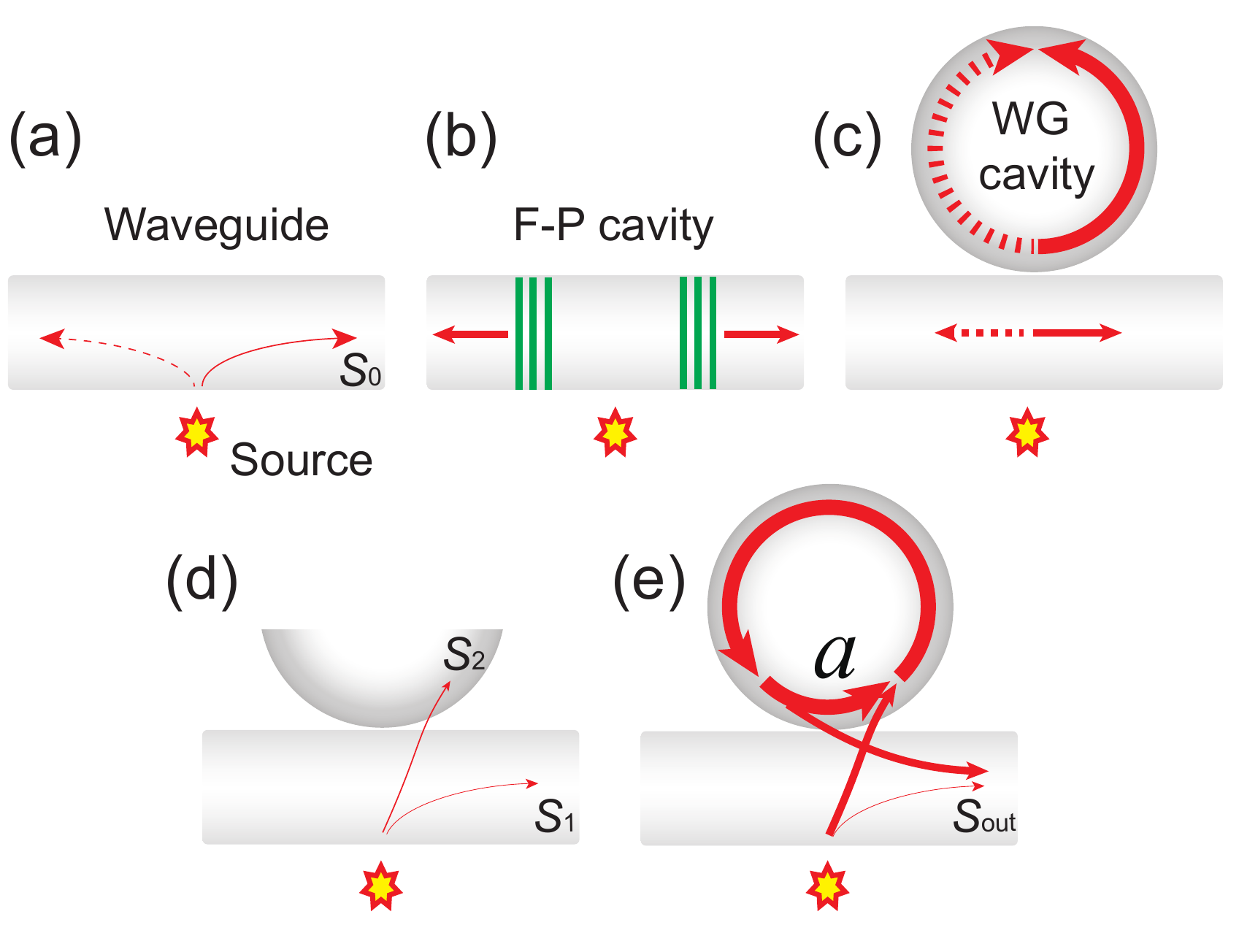}
\caption{Schematics of whispering gallery (WG) cavity-enhanced directional coupling between a light source and a waveguide. (a)~Without a cavity, the light is coupled to a waveguide mode with an amplitude, $s_0$. (b) The coupling efficiency can be enhanced by a F-P cavity resonance, but directionality is lost. (c) A WG cavity can enhance the coupling efficiency and maintain the directionality. (d)~The cavity introduces scattering losses at the contact point, while, at the same time, the light is coupled into the (cavity-free) guided mode with an amplitude, $s_2$. (e)~When the resonance condition is satisfied, the output mode amplitude, $s_{out}$, is enhanced by a WG mode.}
\label{fig:fig1}
\end{figure}

For the case of \textit{emitters}, it was found that the coupling efficiency reaches about 28\% for atoms on the surface of a single-mode nanofiber~\cite{nayak_njp_2008} and 83\% for quantum dots coupled to a nanoantenna~\cite{curto2010unidirectional}. However, the total $\eta$ values remain very low, due to the small size of the secondary source (that is the emitter). This holds for any geometry with direct, non-resonant coupling between the light source and the waveguide, see Fig.~\ref{fig:fig1}(a). To the best of our knowledge, the most efficient directional coupler with this configuration is the recently demonstrated cross-fiber coupler (with the source being the scattering from the fiber crossing point) where values of $\eta\sim10^{-3}$ were achieved~\cite{PhysRevApplied.11.064041}.

By introducing a resonator into the system, one may expect an enhancement of the coupling efficiency. Indeed, it was demonstrated that up to 94\% (instead of 28\%) of spontaneous emission from atoms can be transformed into the guided modes of a nanofiber by using two Bragg-grating mirrors to form a Fabry-P{\'e}rot cavity (F-P in Fig. \ref{fig:fig1}(b)) around the ultrathin waist region where the atoms interact with the optical modes~\cite{le2009cavity}. However, such enhancement occurs due to interference between multiply reflected modes, which have to be quasi-linearly polarized. Consequently, the emission is coupled equally into modes that are guided towards either end of the waveguide and the coupler loses its directionality.

In order to enhance coupling and keep it directional, it is logical to apply a directional resonator, such as a whispering gallery (WG) cavity. Such cavities support pairs of degenerate modes, circulating in opposite directions along loop trajectories. The application of WG cavities for enhanced coupling of light from single emitters has been proposed  previously (see~\cite{lodahl2017chiral} and references therein, also~\cite{martin2019chiral}). However, the cavity alone does not resolve the issue of low total~$\eta$ values. In this work, we suggest a different approach to this problem: instead of coupling a chiral light source directly to a WG cavity, we place an output waveguide between the cavity and the source, as sketched in Fig.~\ref{fig:fig1}(c). In this configuration, the source (input fiber) does not hinder the Q-factors of the WG modes by scattering, and the output modes are enhanced due to the high optical state density caused by the resonance. This strategy enables us to achieve an order of magnitude enhancement of the directional coupling between two crossed optical nanofibers, thus reaching $\eta\sim10^{-2}$, with some room for further improvement. Since the source  is not in direct contact with the cavity (in this work, a silica microsphere), the latter maintains a high Q-factor. Moreover, we found that both transverse magnetic (TM) and transverse electric (TE) cavity modes experience the enhancement. Previously, only TM modes were considered to have the potential for enhancing the directional coupling between a light source and a WG cavity, due to rolling of the electric field~\cite{junge2013strong}.

Let us first consider the mechanism for resonantly enhanced directional coupling. In the absence of the WG cavity (Fig.~\ref{fig:fig1}(a)), a small portion of light from the source can be coupled to an output waveguide mode with an amplitude, $s_0$. When a microsphere (non-resonant) is brought in contact with the waveguide (Fig.~\ref{fig:fig1}(d)), scattering losses are introduced by radiation modes, and the light is partially coupled to a cavity-free guided mode with an amplitude, $s_2$; hence, the output signal has a reduced amplitude, $s_1<s_0$. Under the resonant condition (Fig.~\ref{fig:fig1}(e)), a WG mode can extract more power from the source and enhance the output signal: $s_{out}>s_0$. We define the amplitude coefficients, $\beta=s_1/s_0$ and $\gamma_{int}=s_2/s_0$ (both much smaller than unity, in general), and express the cavity mode amplitude, $a$, as follows~\cite{gorodetsky1999optical}:
\begin{equation}
    a(t)=Ta(t-\tau_0){\rm exp}({-\alpha L/2+i2\pi n_{eff}L/\lambda})+\gamma_{int} s_0,
    \label{eq:eq1}
\end{equation}
where $T$ is the transmission coefficient of the cavity at the coupling region, $\tau_0=n_{eff}L/c$ is the circulation time for the WG mode, $L$ is the circumference of the microsphere, $\lambda$ is the wavelength, $n_{eff}$ is the effective refractive index for the mode, $c$ is the speed of light in vacuum, and $\alpha$  is the linear attenuation due to the intrinsic loss. 

According to Eq. (\ref{eq:eq1}), we can write the mode amplitude in the steady-state:
\begin{equation}
    A=\frac{\gamma_{int} s_0}{(\kappa_0+\kappa_{ext}+i\Delta\omega)\tau_0},
\end{equation}
where $\kappa_0=\alpha c/(2n_{eff})$, $\kappa_{ext}=(1-T)/\tau_0$, and $\Delta\omega$ is the detuning. Therefore, the amplitude of the output field is
\begin{equation}
    s_{out}=\sqrt{2\kappa_{ext}\tau_0}A+\beta s_0.
\end{equation}
Note that, near a resonant frequency, $\beta s_0\ll \sqrt{2\kappa_{ext}\tau_0} A$, $\beta s_0$ can be neglected, and we obtain
\begin{equation}
    \frac{s_{out}}{s_0}=\frac{\gamma_{int}\sqrt{2\kappa_{ext}}}{(\kappa_0+\kappa_{ext}+i\Delta\omega)\sqrt{\tau_0}}\,.
\end{equation}
The above derivation only exploits the directional resonant property of the WG cavity.  This result is universal and should not be restricted to a certain cavity mode, such as TE or TM, as long as there exists nonzero coupling between the waveguide and the cavity mode. For the resonant case, the enhancement is ${\cal E}=|s_{out}|^2/|s_0|^2\propto ({Q}/{Q_{ext})(Q/L)}$, where $Q$ and $Q_{ext}$ are the total and external quality factors, respectively. It is similar in construction to the Purcell enhancement factor, $F\propto Q/V$~\cite{kippenberg2009purcell,novotny2012principles,martin2019chiral}. It is worth noting that the WG resonators cannot be considered to be \textit{directional}, in some cases, where the two original, degenerate, counter-propagating traveling modes are strongly coupled due to  backscattering \cite{zhu2010chip}. In such scenarios, the new eigenmodes are pairs of split standing-wave modes and, in particular if the splittings are larger than the resonance linewidths, the WG resonators behave similarly to F-P resonators. To avoid mode splitting and to maintain high Q values, a WG resonator with a smooth surface is preferred in the experiment.

\begin{figure}[t]
\centering
\includegraphics[width=1\linewidth]{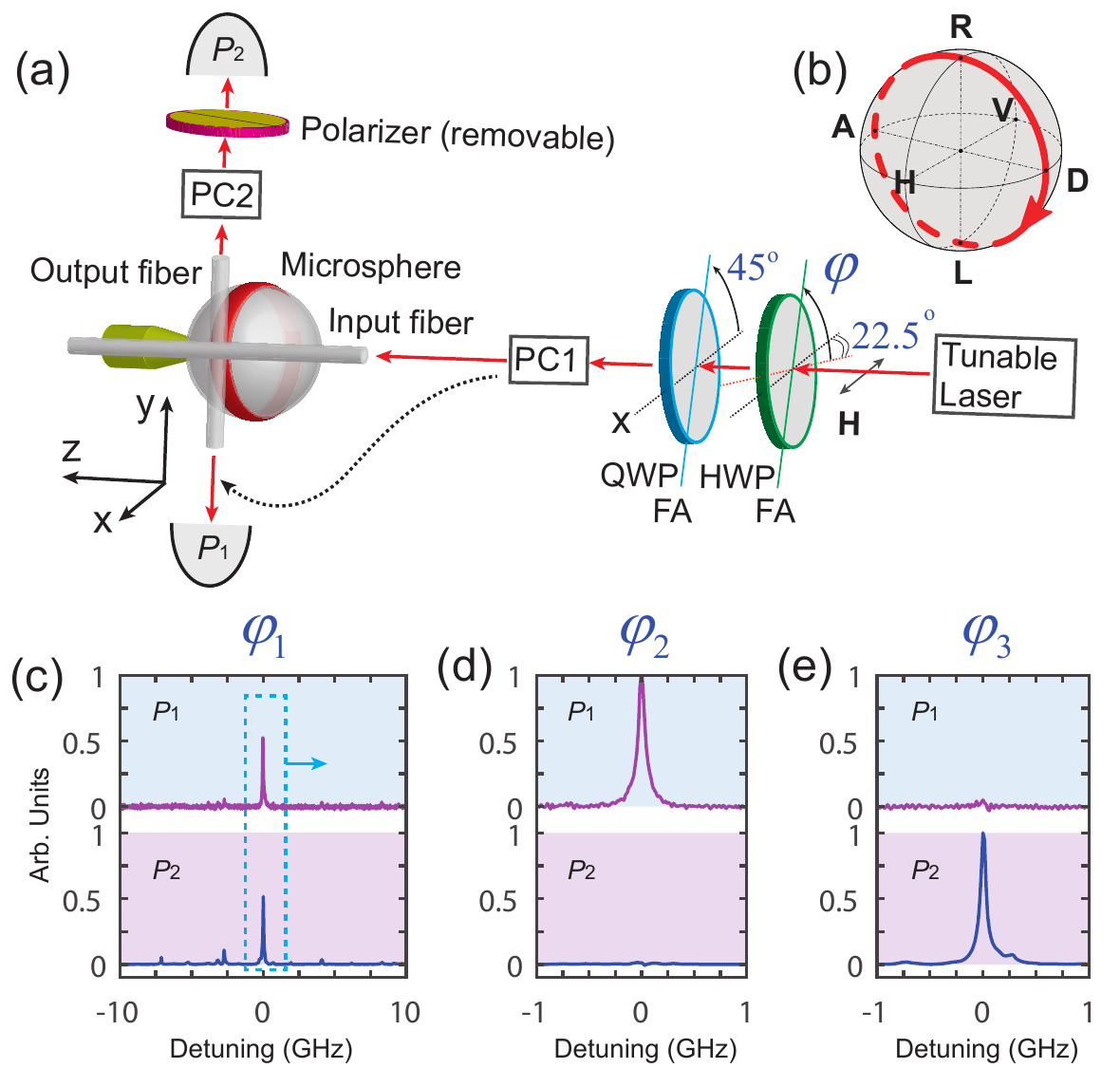}
\caption{(a) Experimental setup, where the directional coupler is represented by two optical nanofibers (input and output) crossed at right angles. Directionality manifests itself in generally unequal output power values, $P_1$ and $P_2$, dependent on the polarization state of the input laser beam, controlled by a  half-wave and a quarter-wave plate (HWP, QWP). Unknown polarization transformations in the fibers are reversed by compensators, PC1 and PC2 (the latter is adjusted with the laser beam sent into the output fiber, see the dotted arrow). By placing a microsphere cavity in contact with the output fiber behind the fiber crossing point, we achieve resonant enhancement of the output signals by coupling to WG modes. TE and TM cavity modes are distinguished by means of a linear polarizer. (b) The input polarization state traces the $\bf DLAR$ circle  on the Poincar{\'e} sphere with varying orientation of the HWP. (c)~Measured spectra of $P_1$ and $P_2$ centered on the strongest peak.  The Q of this cavity mode is around $10^6$. (d),(e)~Zoomed-in spectral peaks at different polarization states defined by the HWP orientation,~$\varphi$. }
\label{fig:fig2}
\end{figure}

Our experimental setup, shown in Fig.~\ref{fig:fig2}(a), is enclosed in a chamber under ambient conditions. The cylindrical waists of two tapered fibers are crossed at right angles to form a directional coupler~\cite{ward_rsi_2014}. The Cartesian coordinate system, $(x,y,z)$, originates in the middle of the input fiber waist and $z$ is parallel to this fiber's axis. The waist diameters of 420~nm~were chosen in order to achieve a high coupling efficiency at the working wavelength of 980~nm~\cite{PhysRevApplied.11.064041}.
In order to enhance the directional coupling between the crossed nanofibers, we use a silica microsphere (with a diameter $\approx 120 \mu$m) touching the output fiber at the point lying on the $x$ axis.
The microsphere is formed by reflowing the tip of a tapered optical fiber using a CO$_2$ laser. To utilize the WG modes (red ring in Fig.~\ref{fig:fig2}(a)) in the $xy$ plane, the microsphere's stem is oriented along the $z$ axis. A collimated Gaussian beam from a tunable laser is launched into the input fiber pigtail after passing through polarization elements, namely a half-wave and a quarter-wave plate (HWP, QWP), 
for generation of the desired polarization states in free-space, and a compensator (PC1, consisting of two QWPs and a variable retarder~\cite{PhysRevApplied.11.064041}) for translation of the generated state to the waist region of the input fiber. With the fast axis (FA) of the QWP being fixed at $45^{\circ}$ with respect to the $-x$ axis, and the HWP having its FA oriented at a variable angle, $({\varphi}-22.5^{\circ})$, the input polarization state travels on the Poincar{\'e} sphere along the circle passing through the diagonal ($\bf D$, linear at $45^{\circ}$ to $x$), left-handed circular ($\bf L$), anti-diagonal ($\bf A$, at $-45^{\circ}$), and right-handed circular ($\bf R$) states, see Fig. \ref{fig:fig2}(b). The power values at the two ends of the output fiber, $P_1$ and $P_2$, are measured with amplified Si-based photodetectors. For circular trajectories concentric with the Poincar{\'e} sphere, these values follow a sinusoidal dependence: $P_1\propto \sin^2[2(\varphi+\varphi_0)]$ and  $P_2\propto \sin^2[2(\varphi+\varphi_0+\Delta \varphi)]$, where $\varphi_0$ and $\Delta \varphi$ are constants defined by the choice of the circle~\cite{PhysRevApplied.11.064041}.  In the case of the $\bf DLAR$ circle, the phase difference, $\Delta \varphi$, is close to the maximum value of $45^{\circ}$, corresponding to the complete directionality when the maximum of $P_1$ occurs at the minimum of $P_2$, and vice versa.

With the laser frequency scanned in the range of 50~GHz, we measure the spectra of $P_1$ and $P_2$ for every polarization state. The spectra depend on the position of the microsphere; therefore, it is kept fixed while acquiring any complete data set. The spectra normally show several Lorentzian peaks, with one being the most prominent, see Fig.~\ref{fig:fig2}(c). Obviously, the peaks correspond to the resonant condition. This resonantly enhanced coupler can demonstrate almost complete directional output: for the HWP set at $\varphi_1$, the peak power values are equal (Fig.~\ref{fig:fig2}(c)); for $\varphi_2\approx\varphi_1-19^{\circ}$, the peak $P_1$ is nearly at a maximum whereas $P_2$ is close to zero (Fig.~\ref{fig:fig2}(d)); and for $\varphi_3\approx\varphi_1+20^{\circ}$, the situation is reversed (Fig.~\ref{fig:fig2}(e)).

By monitoring the output polarization states from the output fiber, we can determine which type of WG modes (TE or TM) contributes to the enhancement. For this purpose, we use another compensator (PC2 in Fig.~\ref{fig:fig2}(a)) and a linear polarizer at one end of the output fiber. Before introducing the cavity, we send the laser beam into the \textit{output} fiber along $y>0$ (see the dotted arrow in Fig.~\ref{fig:fig2}(a)), set the field at the waist to be $z$-polarized (by means of PC1 and the power readings from the \textit{input} fiber), and adjust PC2 to reach the maximum $P_2$ with the polarizer's transmission axis parallel to~$z$. Once this is done, the WG modes can be identified by checking whether the maximum $P_2$ corresponds to the transmission axis parallel to $x$ (TM mode) or $z$ (TE mode). Tilted output polarizations indicate non-equatorial (that is, not lying in the $xy$ plane) WG modes. These may occur if the microsphere and the crossed fibers are misaligned. We avoided such situations because they usually resulted in weaker or less-balanced output signals.

\begin{figure}[t]
\centering
\includegraphics[width=1\linewidth]{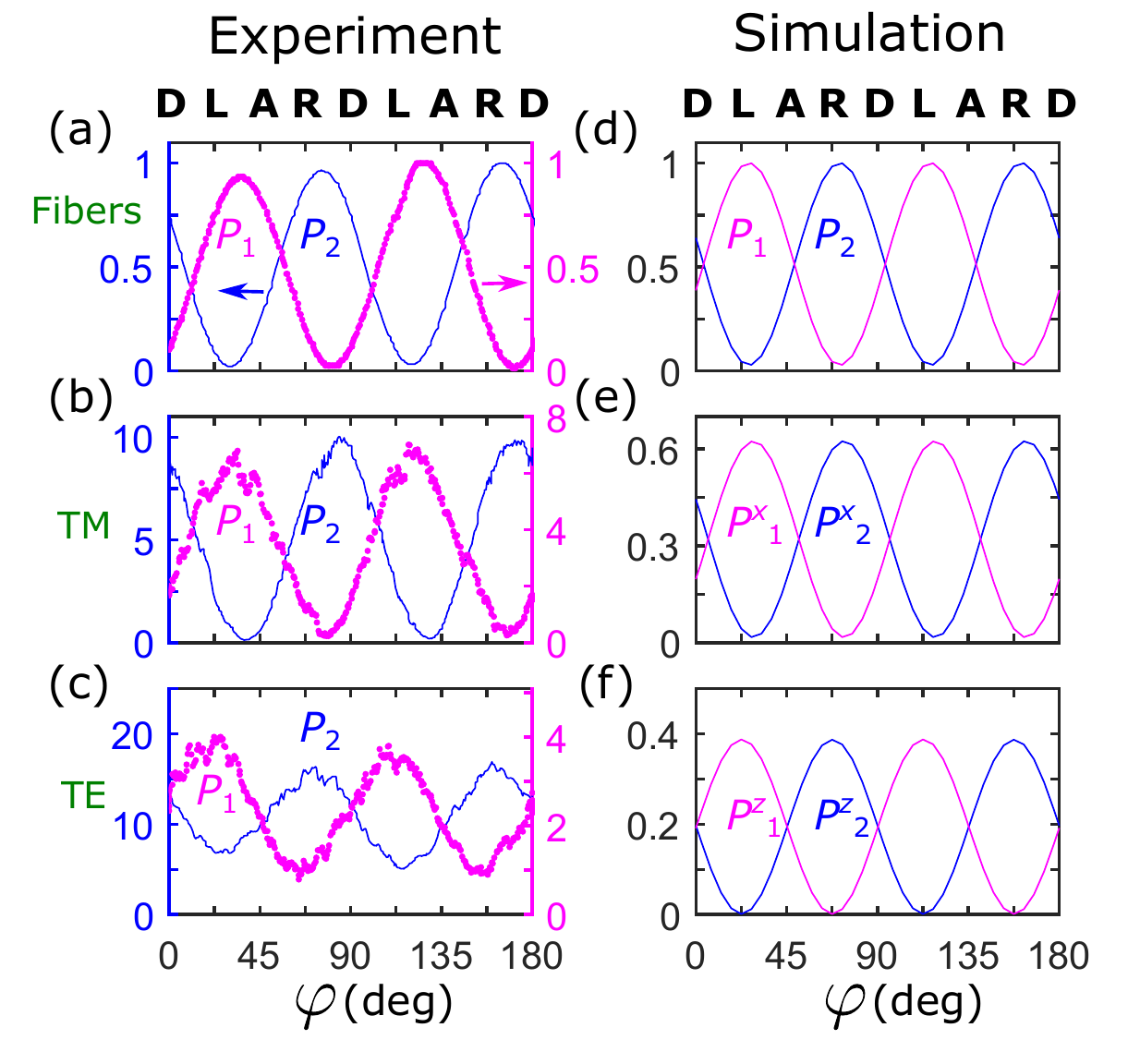}
\caption{Experimental (left) and simulated (right) operation of the directional cross-fiber coupler. (a),(d)~Output power values, $P_1$ and $P_2$, measured without a cavity. (b),(c)~Same, measured in the presence of the microsphere, for the case of enhancement with TM or TE WG modes, respectively. (e),(f)~Simulated power values for $x$- and $z$-polarized fundamental modes of the output fiber  without a cavity, respectively. All the plots are normalized to the maxima of $P_1$ and $P_2$ obtained without a cavity.}
\label{fig:fig3}
\end{figure}

A sample polarization dependent response of the cavity-enhanced directional coupler is presented in the left side of Fig.~\ref{fig:fig3}, where the peak power values in a selected mode are plotted as a function of the HWP orientation,~$\varphi$. As a reference, we initially characterized the crossed-fiber coupler without the microsphere, see Fig.~\ref{fig:fig3}(a), where the measured power values are normalized to the maximum. The measured maximum coupling efficiency is $\eta_0\approx0.15$\%. With the addition of the microsphere, an order of magnitude increase to the output signals is seen for both TM and TE modes (Figs.~\ref{fig:fig3}(b),(c)). The observed imbalance of the output (the maximum $P_2$ being larger than the maximum $P_1$) was reproducible over multiple experimental attempts. 
We attribute this systematic error to some structural asymmetries in the setup, e.~g., the two-fiber crossing point being misaligned with respect to the fiber-microsphere contact point. However, it is quite challenging to  tune the coupling continuously during experiments due to the static force. Therefore, we randomly set the coupling and record different resonances by tuning the laser wavelength in the experiment. The data shown in Fig. ~\ref{fig:fig3}(b)(c) only represent one specific example. In general, we do not find any obvious difference between TM and TE modes with regards to  the magnitude of the enhancement. We expect that if the coupler were built on a chip using flat WG microcavities, such as  discs or toroids, the directional output would be more symmetrical  because of their simpler geometries. 
Notably, when the microsphere was placed \textit{between} the two nanofibers, no directional coupling signal was detected. 
 We attribute this result to the unmediated coupling between the input fiber and the perpendicular WG modes being too weak compared to that between two crossed fibers. The scheme with a WG cavity between the source and the output fiber is still viable if the source is a small nanoparticle\cite{ward2019excitation,shu2018scatterer} or an atom~\cite{kippenberg2009purcell,junge2013strong,zhu2014interfacing,martin2019chiral} instead of a nanofiber which may interact with the cavity over a relatively large and elongated area.

To explain the phenomenon that the directional coupling can be enhanced by both TM and TE modes, it is worth revisiting the mechanism of directional coupling for the simpler cross-fiber system, which we simulate using a finite element method, see the results on the right-hand side of Fig.~\ref{fig:fig3}. In this light-matter system, the mirror symmetry can be broken not only by the spin momentum locking effect, but also by a tilted linear polarization~\cite{PhysRevApplied.11.064041}, a phenomenon which is not expected for a point-like emitter scenario~\cite{petersen2014chiral}. Indeed, in the cross-fiber coupler,  light guided by the input fiber can be directionally coupled to both the $HE_{11}^x$ and $HE_{11}^z$ fundamental modes of the output fiber with similar amplitudes, see the corresponding simulated power values in Figs.~\ref{fig:fig3}(e),(f). Consequently, both TM and TE modes of the microsphere can be excited when it is coupled to the output nanofiber.

In our system, the enhancement, $\cal E$, can be attributed to the maintained high-Q value and the small size of the cavity. However, $\cal E$ is still limited to the order of 10, because of the small coupling coefficient, $\gamma_{int}$. We found that with a thinner output fiber, $\cal E$ can exceed 100. However, in this case, $|s_0|^2$ also decreases dramatically, and the total coupling efficiency, $\eta=\eta_0{\cal E}$, is still limited to a few percent. We believe that it can be further improved by redesigning the WG cavity. For example, the solid microsphere could be replaced by a hollow, thin-walled microbubble cavity~\cite{Ward:18}. Since the latter has a lower effective refractive index, coupling with the waveguide could be improved and scattering losses reduced.
 
In conclusion, we have presented an experimental demonstration of cavity-enhanced directional coupling between two crossed nanofiber waveguides. A microsphere whispering gallery resonator brought in contact with the output waveguide can reliably increase the directional signal from the coupler by an order of magnitude. As confirmed by polarization analysis, both transverse magnetic and transverse electric modes of the cavity can contribute to the enhancement. Our results suggest that higher coupling efficiencies could be achieved when light sources in cavity-based directional nanophotonic circuits are placed on the output waveguide, instead of being brought into direct contact with the cavity.

\noindent {\bf Funding.} This work was funded by the Okinawa Institute of Science and Technology Graduate University (OIST). G.~T.~was supported by the Japan Society for the Promotion of Science (JSPS) as an International Research Fellow (Standard, ID~No.P18367).

\bibliography{reff}

\begin{thebibliography}{32}%
\makeatletter
\providecommand \@ifxundefined [1]{%
 \@ifx{#1\undefined}
}%
\providecommand \@ifnum [1]{%
 \ifnum #1\expandafter \@firstoftwo
 \else \expandafter \@secondoftwo
 \fi
}%
\providecommand \@ifx [1]{%
 \ifx #1\expandafter \@firstoftwo
 \else \expandafter \@secondoftwo
 \fi
}%
\providecommand \natexlab [1]{#1}%
\providecommand \enquote  [1]{``#1''}%
\providecommand \bibnamefont  [1]{#1}%
\providecommand \bibfnamefont [1]{#1}%
\providecommand \citenamefont [1]{#1}%
\providecommand \href@noop [0]{\@secondoftwo}%
\providecommand \href [0]{\begingroup \@sanitize@url \@href}%
\providecommand \@href[1]{\@@startlink{#1}\@@href}%
\providecommand \@@href[1]{\endgroup#1\@@endlink}%
\providecommand \@sanitize@url [0]{\catcode `\\12\catcode `\$12\catcode
  `\&12\catcode `\#12\catcode `\^12\catcode `\_12\catcode `\%12\relax}%
\providecommand \@@startlink[1]{}%
\providecommand \@@endlink[0]{}%
\providecommand \url  [0]{\begingroup\@sanitize@url \@url }%
\providecommand \@url [1]{\endgroup\@href {#1}{\urlprefix }}%
\providecommand \urlprefix  [0]{URL }%
\providecommand \Eprint [0]{\href }%
\providecommand \doibase [0]{http://dx.doi.org/}%
\providecommand \selectlanguage [0]{\@gobble}%
\providecommand \bibinfo  [0]{\@secondoftwo}%
\providecommand \bibfield  [0]{\@secondoftwo}%
\providecommand \translation [1]{[#1]}%
\providecommand \BibitemOpen [0]{}%
\providecommand \bibitemStop [0]{}%
\providecommand \bibitemNoStop [0]{.\EOS\space}%
\providecommand \EOS [0]{\spacefactor3000\relax}%
\providecommand \BibitemShut  [1]{\csname bibitem#1\endcsname}%
\let\auto@bib@innerbib\@empty
%</preamble>
\bibitem [{\citenamefont {Novotny}\ and\ \citenamefont
  {Hecht}(2012)}]{novotny2012principles}%
  \BibitemOpen
  \bibfield  {author} {\bibinfo {author} {\bibfnamefont {L.}~\bibnamefont
  {Novotny}}\ and\ \bibinfo {author} {\bibfnamefont {B.}~\bibnamefont
  {Hecht}},\ }\href@noop {} {\emph {\bibinfo {title} {Principles of
  nano-optics}}}\ (\bibinfo  {publisher} {Cambridge University Press},\
  \bibinfo {year} {2012})\BibitemShut {NoStop}%
\bibitem [{\citenamefont {Noda}, \citenamefont {Fujita},\ and\ \citenamefont
  {Asano}(2007)}]{noda2007spontaneous}%
  \BibitemOpen
  \bibfield  {author} {\bibinfo {author} {\bibfnamefont {S.}~\bibnamefont
  {Noda}}, \bibinfo {author} {\bibfnamefont {M.}~\bibnamefont {Fujita}}, \ and\
  \bibinfo {author} {\bibfnamefont {T.}~\bibnamefont {Asano}},\ }\href@noop {}
  {\bibfield  {journal} {\bibinfo  {journal} {Nat. Photonics}\ }\textbf
  {\bibinfo {volume} {1}},\ \bibinfo {pages} {449} (\bibinfo {year}
  {2007})}\BibitemShut {NoStop}%
\bibitem [{\citenamefont {Kippenberg}\ \emph {et~al.}(2009)\citenamefont
  {Kippenberg}, \citenamefont {Tchebotareva}, \citenamefont {Kalkman},
  \citenamefont {Polman},\ and\ \citenamefont
  {Vahala}}]{kippenberg2009purcell}%
  \BibitemOpen
  \bibfield  {author} {\bibinfo {author} {\bibfnamefont {T.~J.}\ \bibnamefont
  {Kippenberg}}, \bibinfo {author} {\bibfnamefont {A.}~\bibnamefont
  {Tchebotareva}}, \bibinfo {author} {\bibfnamefont {J.}~\bibnamefont
  {Kalkman}}, \bibinfo {author} {\bibfnamefont {A.}~\bibnamefont {Polman}}, \
  and\ \bibinfo {author} {\bibfnamefont {K.~J.}\ \bibnamefont {Vahala}},\
  }\href@noop {} {\bibfield  {journal} {\bibinfo  {journal} {Phys. Rev. Lett.}\
  }\textbf {\bibinfo {volume} {103}},\ \bibinfo {pages} {027406} (\bibinfo
  {year} {2009})}\BibitemShut {NoStop}%
\bibitem [{\citenamefont {Zhu}\ \emph {et~al.}(2014)\citenamefont {Zhu},
  \citenamefont {{\"O}zdemir}, \citenamefont {Yilmaz}, \citenamefont {Peng},
  \citenamefont {Dong}, \citenamefont {Tomes}, \citenamefont {Carmon},\ and\
  \citenamefont {Yang}}]{zhu2014interfacing}%
  \BibitemOpen
  \bibfield  {author} {\bibinfo {author} {\bibfnamefont {J.}~\bibnamefont
  {Zhu}}, \bibinfo {author} {\bibfnamefont {{\c{S}}.~K.}\ \bibnamefont
  {{\"O}zdemir}}, \bibinfo {author} {\bibfnamefont {H.}~\bibnamefont {Yilmaz}},
  \bibinfo {author} {\bibfnamefont {B.}~\bibnamefont {Peng}}, \bibinfo {author}
  {\bibfnamefont {M.}~\bibnamefont {Dong}}, \bibinfo {author} {\bibfnamefont
  {M.}~\bibnamefont {Tomes}}, \bibinfo {author} {\bibfnamefont
  {T.}~\bibnamefont {Carmon}}, \ and\ \bibinfo {author} {\bibfnamefont
  {L.}~\bibnamefont {Yang}},\ }\href@noop {} {\bibfield  {journal} {\bibinfo
  {journal} {Sci. Rep.}\ }\textbf {\bibinfo {volume} {4}},\ \bibinfo {pages}
  {6396} (\bibinfo {year} {2014})}\BibitemShut {NoStop}%
\bibitem [{\citenamefont {Ward}\ \emph {et~al.}(2019)\citenamefont {Ward},
  \citenamefont {Lei}, \citenamefont {Vincent}, \citenamefont {Gupta},
  \citenamefont {Mondal}, \citenamefont {Fick},\ and\ \citenamefont {{Nic
  Chormaic}}}]{ward2019excitation}%
  \BibitemOpen
  \bibfield  {author} {\bibinfo {author} {\bibfnamefont {J.~M.}\ \bibnamefont
  {Ward}}, \bibinfo {author} {\bibfnamefont {F.}~\bibnamefont {Lei}}, \bibinfo
  {author} {\bibfnamefont {S.}~\bibnamefont {Vincent}}, \bibinfo {author}
  {\bibfnamefont {P.}~\bibnamefont {Gupta}}, \bibinfo {author} {\bibfnamefont
  {S.~K.}\ \bibnamefont {Mondal}}, \bibinfo {author} {\bibfnamefont
  {J.}~\bibnamefont {Fick}}, \ and\ \bibinfo {author} {\bibfnamefont
  {S.}~\bibnamefont {{Nic Chormaic}}},\ }\href@noop {} {\bibfield  {journal}
  {\bibinfo  {journal} {Opt. Lett.}\ }\textbf {\bibinfo {volume} {44}},\
  \bibinfo {pages} {3386} (\bibinfo {year} {2019})}\BibitemShut {NoStop}%
\bibitem [{\citenamefont {Curto}\ \emph {et~al.}(2010)\citenamefont {Curto},
  \citenamefont {Volpe}, \citenamefont {Taminiau}, \citenamefont {Kreuzer},
  \citenamefont {Quidant},\ and\ \citenamefont {van
  Hulst}}]{curto2010unidirectional}%
  \BibitemOpen
  \bibfield  {author} {\bibinfo {author} {\bibfnamefont {A.~G.}\ \bibnamefont
  {Curto}}, \bibinfo {author} {\bibfnamefont {G.}~\bibnamefont {Volpe}},
  \bibinfo {author} {\bibfnamefont {T.~H.}\ \bibnamefont {Taminiau}}, \bibinfo
  {author} {\bibfnamefont {M.~P.}\ \bibnamefont {Kreuzer}}, \bibinfo {author}
  {\bibfnamefont {R.}~\bibnamefont {Quidant}}, \ and\ \bibinfo {author}
  {\bibfnamefont {N.~F.}\ \bibnamefont {van Hulst}},\ }\href@noop {} {\bibfield
   {journal} {\bibinfo  {journal} {Science}\ }\textbf {\bibinfo {volume}
  {329}},\ \bibinfo {pages} {930} (\bibinfo {year} {2010})}\BibitemShut
  {NoStop}%
\bibitem [{\citenamefont {Lee}\ \emph {et~al.}(2011)\citenamefont {Lee},
  \citenamefont {Chen}, \citenamefont {Eghlidi}, \citenamefont {Kukura},
  \citenamefont {Lettow}, \citenamefont {Renn}, \citenamefont {Sandoghdar},\
  and\ \citenamefont {G{\"o}tzinger}}]{lee2011planar}%
  \BibitemOpen
  \bibfield  {author} {\bibinfo {author} {\bibfnamefont {K.~G.}\ \bibnamefont
  {Lee}}, \bibinfo {author} {\bibfnamefont {X.}~\bibnamefont {Chen}}, \bibinfo
  {author} {\bibfnamefont {H.}~\bibnamefont {Eghlidi}}, \bibinfo {author}
  {\bibfnamefont {P.}~\bibnamefont {Kukura}}, \bibinfo {author} {\bibfnamefont
  {R.}~\bibnamefont {Lettow}}, \bibinfo {author} {\bibfnamefont
  {A.}~\bibnamefont {Renn}}, \bibinfo {author} {\bibfnamefont {V.}~\bibnamefont
  {Sandoghdar}}, \ and\ \bibinfo {author} {\bibfnamefont {S.}~\bibnamefont
  {G{\"o}tzinger}},\ }\href@noop {} {\bibfield  {journal} {\bibinfo  {journal}
  {Nat. Photonics}\ }\textbf {\bibinfo {volume} {5}},\ \bibinfo {pages} {166}
  (\bibinfo {year} {2011})}\BibitemShut {NoStop}%
\bibitem [{\citenamefont {Bliokh}\ \emph {et~al.}(2015)\citenamefont {Bliokh},
  \citenamefont {Rodr{\'\i}guez-Fortu{\~n}o}, \citenamefont {Nori},\ and\
  \citenamefont {Zayats}}]{bliokh2015spin}%
  \BibitemOpen
  \bibfield  {author} {\bibinfo {author} {\bibfnamefont {K.~Y.}\ \bibnamefont
  {Bliokh}}, \bibinfo {author} {\bibfnamefont {F.~J.}\ \bibnamefont
  {Rodr{\'\i}guez-Fortu{\~n}o}}, \bibinfo {author} {\bibfnamefont
  {F.}~\bibnamefont {Nori}}, \ and\ \bibinfo {author} {\bibfnamefont {A.~V.}\
  \bibnamefont {Zayats}},\ }\href@noop {} {\bibfield  {journal} {\bibinfo
  {journal} {Nat. Photonics}\ }\textbf {\bibinfo {volume} {9}},\ \bibinfo
  {pages} {796} (\bibinfo {year} {2015})}\BibitemShut {NoStop}%
\bibitem [{\citenamefont {Lee}\ \emph {et~al.}(2012)\citenamefont {Lee},
  \citenamefont {Lee}, \citenamefont {Park}, \citenamefont {Oh}, \citenamefont
  {Lee}, \citenamefont {Kim},\ and\ \citenamefont {Lee}}]{lee2012role}%
  \BibitemOpen
  \bibfield  {author} {\bibinfo {author} {\bibfnamefont {S.-Y.}\ \bibnamefont
  {Lee}}, \bibinfo {author} {\bibfnamefont {I.-M.}\ \bibnamefont {Lee}},
  \bibinfo {author} {\bibfnamefont {J.}~\bibnamefont {Park}}, \bibinfo {author}
  {\bibfnamefont {S.}~\bibnamefont {Oh}}, \bibinfo {author} {\bibfnamefont
  {W.}~\bibnamefont {Lee}}, \bibinfo {author} {\bibfnamefont {K.-Y.}\
  \bibnamefont {Kim}}, \ and\ \bibinfo {author} {\bibfnamefont
  {B.}~\bibnamefont {Lee}},\ }\href@noop {} {\bibfield  {journal} {\bibinfo
  {journal} {Phys. Rev. Lett.}\ }\textbf {\bibinfo {volume} {108}},\ \bibinfo
  {pages} {213907} (\bibinfo {year} {2012})}\BibitemShut {NoStop}%
\bibitem [{\citenamefont {Lin}\ \emph {et~al.}(2013)\citenamefont {Lin},
  \citenamefont {Mueller}, \citenamefont {Wang}, \citenamefont {Yuan},
  \citenamefont {Antoniou}, \citenamefont {Yuan},\ and\ \citenamefont
  {Capasso}}]{lin2013polarization}%
  \BibitemOpen
  \bibfield  {author} {\bibinfo {author} {\bibfnamefont {J.}~\bibnamefont
  {Lin}}, \bibinfo {author} {\bibfnamefont {J.~B.}\ \bibnamefont {Mueller}},
  \bibinfo {author} {\bibfnamefont {Q.}~\bibnamefont {Wang}}, \bibinfo {author}
  {\bibfnamefont {G.}~\bibnamefont {Yuan}}, \bibinfo {author} {\bibfnamefont
  {N.}~\bibnamefont {Antoniou}}, \bibinfo {author} {\bibfnamefont {X.-C.}\
  \bibnamefont {Yuan}}, \ and\ \bibinfo {author} {\bibfnamefont
  {F.}~\bibnamefont {Capasso}},\ }\href@noop {} {\bibfield  {journal} {\bibinfo
   {journal} {Science}\ }\textbf {\bibinfo {volume} {340}},\ \bibinfo {pages}
  {331} (\bibinfo {year} {2013})}\BibitemShut {NoStop}%
\bibitem [{\citenamefont {Rodr{\'\i}guez-Fortu{\~n}o}\ \emph
  {et~al.}(2013)\citenamefont {Rodr{\'\i}guez-Fortu{\~n}o}, \citenamefont
  {Marino}, \citenamefont {Ginzburg}, \citenamefont {O’Connor}, \citenamefont
  {Mart{\'\i}nez}, \citenamefont {Wurtz},\ and\ \citenamefont
  {Zayats}}]{rodriguez2013near}%
  \BibitemOpen
  \bibfield  {author} {\bibinfo {author} {\bibfnamefont {F.~J.}\ \bibnamefont
  {Rodr{\'\i}guez-Fortu{\~n}o}}, \bibinfo {author} {\bibfnamefont
  {G.}~\bibnamefont {Marino}}, \bibinfo {author} {\bibfnamefont
  {P.}~\bibnamefont {Ginzburg}}, \bibinfo {author} {\bibfnamefont
  {D.}~\bibnamefont {O’Connor}}, \bibinfo {author} {\bibfnamefont
  {A.}~\bibnamefont {Mart{\'\i}nez}}, \bibinfo {author} {\bibfnamefont {G.~A.}\
  \bibnamefont {Wurtz}}, \ and\ \bibinfo {author} {\bibfnamefont {A.~V.}\
  \bibnamefont {Zayats}},\ }\href@noop {} {\bibfield  {journal} {\bibinfo
  {journal} {Science}\ }\textbf {\bibinfo {volume} {340}},\ \bibinfo {pages}
  {328} (\bibinfo {year} {2013})}\BibitemShut {NoStop}%
\bibitem [{\citenamefont {Luxmoore}\ \emph {et~al.}(2013)\citenamefont
  {Luxmoore}, \citenamefont {Wasley}, \citenamefont {Ramsay}, \citenamefont
  {Thijssen}, \citenamefont {Oulton}, \citenamefont {Hugues}, \citenamefont
  {Kasture}, \citenamefont {Achanta}, \citenamefont {Fox},\ and\ \citenamefont
  {Skolnick}}]{luxmoore2013interfacing}%
  \BibitemOpen
  \bibfield  {author} {\bibinfo {author} {\bibfnamefont {I.}~\bibnamefont
  {Luxmoore}}, \bibinfo {author} {\bibfnamefont {N.}~\bibnamefont {Wasley}},
  \bibinfo {author} {\bibfnamefont {A.}~\bibnamefont {Ramsay}}, \bibinfo
  {author} {\bibfnamefont {A.}~\bibnamefont {Thijssen}}, \bibinfo {author}
  {\bibfnamefont {R.}~\bibnamefont {Oulton}}, \bibinfo {author} {\bibfnamefont
  {M.}~\bibnamefont {Hugues}}, \bibinfo {author} {\bibfnamefont
  {S.}~\bibnamefont {Kasture}}, \bibinfo {author} {\bibfnamefont
  {V.}~\bibnamefont {Achanta}}, \bibinfo {author} {\bibfnamefont
  {A.}~\bibnamefont {Fox}}, \ and\ \bibinfo {author} {\bibfnamefont
  {M.}~\bibnamefont {Skolnick}},\ }\href@noop {} {\bibfield  {journal}
  {\bibinfo  {journal} {Phys. Rev. Lett.}\ }\textbf {\bibinfo {volume} {110}},\
  \bibinfo {pages} {037402} (\bibinfo {year} {2013})}\BibitemShut {NoStop}%
\bibitem [{\citenamefont {Junge}\ \emph {et~al.}(2013)\citenamefont {Junge},
  \citenamefont {O’shea}, \citenamefont {Volz},\ and\ \citenamefont
  {Rauschenbeutel}}]{junge2013strong}%
  \BibitemOpen
  \bibfield  {author} {\bibinfo {author} {\bibfnamefont {C.}~\bibnamefont
  {Junge}}, \bibinfo {author} {\bibfnamefont {D.}~\bibnamefont {O’shea}},
  \bibinfo {author} {\bibfnamefont {J.}~\bibnamefont {Volz}}, \ and\ \bibinfo
  {author} {\bibfnamefont {A.}~\bibnamefont {Rauschenbeutel}},\ }\href@noop {}
  {\bibfield  {journal} {\bibinfo  {journal} {Phys. Rev. Lett.}\ }\textbf
  {\bibinfo {volume} {110}},\ \bibinfo {pages} {213604} (\bibinfo {year}
  {2013})}\BibitemShut {NoStop}%
\bibitem [{\citenamefont {Neugebauer}\ \emph {et~al.}(2014)\citenamefont
  {Neugebauer}, \citenamefont {Bauer}, \citenamefont {Banzer},\ and\
  \citenamefont {Leuchs}}]{neugebauer2014polarization}%
  \BibitemOpen
  \bibfield  {author} {\bibinfo {author} {\bibfnamefont {M.}~\bibnamefont
  {Neugebauer}}, \bibinfo {author} {\bibfnamefont {T.}~\bibnamefont {Bauer}},
  \bibinfo {author} {\bibfnamefont {P.}~\bibnamefont {Banzer}}, \ and\ \bibinfo
  {author} {\bibfnamefont {G.}~\bibnamefont {Leuchs}},\ }\href@noop {}
  {\bibfield  {journal} {\bibinfo  {journal} {Nano Lett.}\ }\textbf {\bibinfo
  {volume} {14}},\ \bibinfo {pages} {2546} (\bibinfo {year}
  {2014})}\BibitemShut {NoStop}%
\bibitem [{\citenamefont {Rodr{\'\i}guez-Fortu{\~n}o}\ \emph
  {et~al.}(2014)\citenamefont {Rodr{\'\i}guez-Fortu{\~n}o}, \citenamefont
  {Barber-Sanz}, \citenamefont {Puerto}, \citenamefont {Griol},\ and\
  \citenamefont {Mart{\'\i}nez}}]{rodriguez2014resolving}%
  \BibitemOpen
  \bibfield  {author} {\bibinfo {author} {\bibfnamefont {F.~J.}\ \bibnamefont
  {Rodr{\'\i}guez-Fortu{\~n}o}}, \bibinfo {author} {\bibfnamefont
  {I.}~\bibnamefont {Barber-Sanz}}, \bibinfo {author} {\bibfnamefont
  {D.}~\bibnamefont {Puerto}}, \bibinfo {author} {\bibfnamefont
  {A.}~\bibnamefont {Griol}}, \ and\ \bibinfo {author} {\bibfnamefont
  {A.}~\bibnamefont {Mart{\'\i}nez}},\ }\href@noop {} {\bibfield  {journal}
  {\bibinfo  {journal} {ACS Photonics}\ }\textbf {\bibinfo {volume} {1}},\
  \bibinfo {pages} {762} (\bibinfo {year} {2014})}\BibitemShut {NoStop}%
\bibitem [{\citenamefont {Petersen}, \citenamefont {Volz},\ and\ \citenamefont
  {Rauschenbeutel}(2014)}]{petersen2014chiral}%
  \BibitemOpen
  \bibfield  {author} {\bibinfo {author} {\bibfnamefont {J.}~\bibnamefont
  {Petersen}}, \bibinfo {author} {\bibfnamefont {J.}~\bibnamefont {Volz}}, \
  and\ \bibinfo {author} {\bibfnamefont {A.}~\bibnamefont {Rauschenbeutel}},\
  }\href@noop {} {\bibfield  {journal} {\bibinfo  {journal} {Science}\ }\textbf
  {\bibinfo {volume} {346}},\ \bibinfo {pages} {67} (\bibinfo {year}
  {2014})}\BibitemShut {NoStop}%
\bibitem [{\citenamefont {Le~Feber}, \citenamefont {Rotenberg},\ and\
  \citenamefont {Kuipers}(2015)}]{le2015nanophotonic}%
  \BibitemOpen
  \bibfield  {author} {\bibinfo {author} {\bibfnamefont {B.}~\bibnamefont
  {Le~Feber}}, \bibinfo {author} {\bibfnamefont {N.}~\bibnamefont {Rotenberg}},
  \ and\ \bibinfo {author} {\bibfnamefont {L.}~\bibnamefont {Kuipers}},\
  }\href@noop {} {\bibfield  {journal} {\bibinfo  {journal} {Nature Commun.}\
  }\textbf {\bibinfo {volume} {6}},\ \bibinfo {pages} {6695} (\bibinfo {year}
  {2015})}\BibitemShut {NoStop}%
\bibitem [{\citenamefont {Shao}\ \emph {et~al.}(2018)\citenamefont {Shao},
  \citenamefont {Zhu}, \citenamefont {Chen}, \citenamefont {Zhang},\ and\
  \citenamefont {Yu}}]{shao2018spin}%
  \BibitemOpen
  \bibfield  {author} {\bibinfo {author} {\bibfnamefont {Z.}~\bibnamefont
  {Shao}}, \bibinfo {author} {\bibfnamefont {J.}~\bibnamefont {Zhu}}, \bibinfo
  {author} {\bibfnamefont {Y.}~\bibnamefont {Chen}}, \bibinfo {author}
  {\bibfnamefont {Y.}~\bibnamefont {Zhang}}, \ and\ \bibinfo {author}
  {\bibfnamefont {S.}~\bibnamefont {Yu}},\ }\href@noop {} {\bibfield  {journal}
  {\bibinfo  {journal} {Nature Commun.}\ }\textbf {\bibinfo {volume} {9}},\
  \bibinfo {pages} {926} (\bibinfo {year} {2018})}\BibitemShut {NoStop}%
\bibitem [{\citenamefont {Lei}\ \emph {et~al.}(2019)\citenamefont {Lei},
  \citenamefont {Tkachenko}, \citenamefont {Ward},\ and\ \citenamefont {{Nic
  Chormaic}}}]{PhysRevApplied.11.064041}%
  \BibitemOpen
  \bibfield  {author} {\bibinfo {author} {\bibfnamefont {F.}~\bibnamefont
  {Lei}}, \bibinfo {author} {\bibfnamefont {G.}~\bibnamefont {Tkachenko}},
  \bibinfo {author} {\bibfnamefont {J.~M.}\ \bibnamefont {Ward}}, \ and\
  \bibinfo {author} {\bibfnamefont {S.}~\bibnamefont {{Nic Chormaic}}},\ }\href
  {\doibase 10.1103/PhysRevApplied.11.064041} {\bibfield  {journal} {\bibinfo
  {journal} {Phys. Rev. Applied}\ }\textbf {\bibinfo {volume} {11}},\ \bibinfo
  {pages} {064041} (\bibinfo {year} {2019})}\BibitemShut {NoStop}%
\bibitem [{\citenamefont {Mitsch}\ \emph {et~al.}(2014)\citenamefont {Mitsch},
  \citenamefont {Sayrin}, \citenamefont {Albrecht}, \citenamefont
  {Schneeweiss},\ and\ \citenamefont {Rauschenbeutel}}]{mitsch2014quantum}%
  \BibitemOpen
  \bibfield  {author} {\bibinfo {author} {\bibfnamefont {R.}~\bibnamefont
  {Mitsch}}, \bibinfo {author} {\bibfnamefont {C.}~\bibnamefont {Sayrin}},
  \bibinfo {author} {\bibfnamefont {B.}~\bibnamefont {Albrecht}}, \bibinfo
  {author} {\bibfnamefont {P.}~\bibnamefont {Schneeweiss}}, \ and\ \bibinfo
  {author} {\bibfnamefont {A.}~\bibnamefont {Rauschenbeutel}},\ }\href@noop {}
  {\bibfield  {journal} {\bibinfo  {journal} {Nature Commun.}\ }\textbf
  {\bibinfo {volume} {5}},\ \bibinfo {pages} {5713} (\bibinfo {year}
  {2014})}\BibitemShut {NoStop}%
\bibitem [{\citenamefont {Shomroni}\ \emph {et~al.}(2014)\citenamefont
  {Shomroni}, \citenamefont {Rosenblum}, \citenamefont {Lovsky}, \citenamefont
  {Bechler}, \citenamefont {Guendelman},\ and\ \citenamefont
  {Dayan}}]{shomroni2014all}%
  \BibitemOpen
  \bibfield  {author} {\bibinfo {author} {\bibfnamefont {I.}~\bibnamefont
  {Shomroni}}, \bibinfo {author} {\bibfnamefont {S.}~\bibnamefont {Rosenblum}},
  \bibinfo {author} {\bibfnamefont {Y.}~\bibnamefont {Lovsky}}, \bibinfo
  {author} {\bibfnamefont {O.}~\bibnamefont {Bechler}}, \bibinfo {author}
  {\bibfnamefont {G.}~\bibnamefont {Guendelman}}, \ and\ \bibinfo {author}
  {\bibfnamefont {B.}~\bibnamefont {Dayan}},\ }\href@noop {} {\bibfield
  {journal} {\bibinfo  {journal} {Science}\ }\textbf {\bibinfo {volume}
  {345}},\ \bibinfo {pages} {903} (\bibinfo {year} {2014})}\BibitemShut
  {NoStop}%
\bibitem [{\citenamefont {S{\"o}llner}\ \emph {et~al.}(2015)\citenamefont
  {S{\"o}llner}, \citenamefont {Mahmoodian}, \citenamefont {Hansen},
  \citenamefont {Midolo}, \citenamefont {Javadi}, \citenamefont
  {Kir{\v{s}}ansk{\.e}}, \citenamefont {Pregnolato}, \citenamefont {El-Ella},
  \citenamefont {Lee}, \citenamefont {Song}, \citenamefont {Stobbe},\ and\
  \citenamefont {Lodahl}}]{sollner2015deterministic}%
  \BibitemOpen
  \bibfield  {author} {\bibinfo {author} {\bibfnamefont {I.}~\bibnamefont
  {S{\"o}llner}}, \bibinfo {author} {\bibfnamefont {S.}~\bibnamefont
  {Mahmoodian}}, \bibinfo {author} {\bibfnamefont {S.~L.}\ \bibnamefont
  {Hansen}}, \bibinfo {author} {\bibfnamefont {L.}~\bibnamefont {Midolo}},
  \bibinfo {author} {\bibfnamefont {A.}~\bibnamefont {Javadi}}, \bibinfo
  {author} {\bibfnamefont {G.}~\bibnamefont {Kir{\v{s}}ansk{\.e}}}, \bibinfo
  {author} {\bibfnamefont {T.}~\bibnamefont {Pregnolato}}, \bibinfo {author}
  {\bibfnamefont {H.}~\bibnamefont {El-Ella}}, \bibinfo {author} {\bibfnamefont
  {E.~H.}\ \bibnamefont {Lee}}, \bibinfo {author} {\bibfnamefont {J.~D.}\
  \bibnamefont {Song}}, \bibinfo {author} {\bibfnamefont {S.}~\bibnamefont
  {Stobbe}}, \ and\ \bibinfo {author} {\bibfnamefont {P.}~\bibnamefont
  {Lodahl}},\ }\href@noop {} {\bibfield  {journal} {\bibinfo  {journal} {Nat.
  Nanotechnol.}\ }\textbf {\bibinfo {volume} {10}},\ \bibinfo {pages} {775}
  (\bibinfo {year} {2015})}\BibitemShut {NoStop}%
\bibitem [{\citenamefont {Coles}\ \emph {et~al.}(2016)\citenamefont {Coles},
  \citenamefont {Price}, \citenamefont {Dixon}, \citenamefont {Royall},
  \citenamefont {Clarke}, \citenamefont {Kok}, \citenamefont {Skolnick},
  \citenamefont {Fox},\ and\ \citenamefont {Makhonin}}]{coles2016chirality}%
  \BibitemOpen
  \bibfield  {author} {\bibinfo {author} {\bibfnamefont {R.}~\bibnamefont
  {Coles}}, \bibinfo {author} {\bibfnamefont {D.}~\bibnamefont {Price}},
  \bibinfo {author} {\bibfnamefont {J.}~\bibnamefont {Dixon}}, \bibinfo
  {author} {\bibfnamefont {B.}~\bibnamefont {Royall}}, \bibinfo {author}
  {\bibfnamefont {E.}~\bibnamefont {Clarke}}, \bibinfo {author} {\bibfnamefont
  {P.}~\bibnamefont {Kok}}, \bibinfo {author} {\bibfnamefont {M.}~\bibnamefont
  {Skolnick}}, \bibinfo {author} {\bibfnamefont {A.}~\bibnamefont {Fox}}, \
  and\ \bibinfo {author} {\bibfnamefont {M.}~\bibnamefont {Makhonin}},\
  }\href@noop {} {\bibfield  {journal} {\bibinfo  {journal} {Nature commun.}\
  }\textbf {\bibinfo {volume} {7}},\ \bibinfo {pages} {11183} (\bibinfo {year}
  {2016})}\BibitemShut {NoStop}%
\bibitem [{\citenamefont {Nayak}\ and\ \citenamefont
  {Hakuta}(2008)}]{nayak_njp_2008}%
  \BibitemOpen
  \bibfield  {author} {\bibinfo {author} {\bibfnamefont {K.~P.}\ \bibnamefont
  {Nayak}}\ and\ \bibinfo {author} {\bibfnamefont {K.}~\bibnamefont {Hakuta}},\
  }\href@noop {} {\bibfield  {journal} {\bibinfo  {journal} {New J. Phys.}\
  }\textbf {\bibinfo {volume} {10}},\ \bibinfo {pages} {053003} (\bibinfo
  {year} {2008})}\BibitemShut {NoStop}%
\bibitem [{\citenamefont {Le~Kien}\ and\ \citenamefont
  {Hakuta}(2009)}]{le2009cavity}%
  \BibitemOpen
  \bibfield  {author} {\bibinfo {author} {\bibfnamefont {F.}~\bibnamefont
  {Le~Kien}}\ and\ \bibinfo {author} {\bibfnamefont {K.}~\bibnamefont
  {Hakuta}},\ }\href@noop {} {\bibfield  {journal} {\bibinfo  {journal} {Phys.
  Rev. A}\ }\textbf {\bibinfo {volume} {80}},\ \bibinfo {pages} {053826}
  (\bibinfo {year} {2009})}\BibitemShut {NoStop}%
\bibitem [{\citenamefont {Lodahl}\ \emph {et~al.}(2017)\citenamefont {Lodahl},
  \citenamefont {Mahmoodian}, \citenamefont {Stobbe}, \citenamefont
  {Rauschenbeutel}, \citenamefont {Schneeweiss}, \citenamefont {Volz},
  \citenamefont {Pichler},\ and\ \citenamefont {Zoller}}]{lodahl2017chiral}%
  \BibitemOpen
  \bibfield  {author} {\bibinfo {author} {\bibfnamefont {P.}~\bibnamefont
  {Lodahl}}, \bibinfo {author} {\bibfnamefont {S.}~\bibnamefont {Mahmoodian}},
  \bibinfo {author} {\bibfnamefont {S.}~\bibnamefont {Stobbe}}, \bibinfo
  {author} {\bibfnamefont {A.}~\bibnamefont {Rauschenbeutel}}, \bibinfo
  {author} {\bibfnamefont {P.}~\bibnamefont {Schneeweiss}}, \bibinfo {author}
  {\bibfnamefont {J.}~\bibnamefont {Volz}}, \bibinfo {author} {\bibfnamefont
  {H.}~\bibnamefont {Pichler}}, \ and\ \bibinfo {author} {\bibfnamefont
  {P.}~\bibnamefont {Zoller}},\ }\href@noop {} {\bibfield  {journal} {\bibinfo
  {journal} {Nature}\ }\textbf {\bibinfo {volume} {541}},\ \bibinfo {pages}
  {473} (\bibinfo {year} {2017})}\BibitemShut {NoStop}%
\bibitem [{\citenamefont {Martin-Cano}, \citenamefont {Haakh},\ and\
  \citenamefont {Rotenberg}(2019)}]{martin2019chiral}%
  \BibitemOpen
  \bibfield  {author} {\bibinfo {author} {\bibfnamefont {D.}~\bibnamefont
  {Martin-Cano}}, \bibinfo {author} {\bibfnamefont {H.~R.}\ \bibnamefont
  {Haakh}}, \ and\ \bibinfo {author} {\bibfnamefont {N.}~\bibnamefont
  {Rotenberg}},\ }\href@noop {} {\bibfield  {journal} {\bibinfo  {journal} {ACS
  Photonics}\ }\textbf {\bibinfo {volume} {6}},\ \bibinfo {pages} {961}
  (\bibinfo {year} {2019})}\BibitemShut {NoStop}%
\bibitem [{\citenamefont {Gorodetsky}\ and\ \citenamefont
  {Ilchenko}(1999)}]{gorodetsky1999optical}%
  \BibitemOpen
  \bibfield  {author} {\bibinfo {author} {\bibfnamefont {M.~L.}\ \bibnamefont
  {Gorodetsky}}\ and\ \bibinfo {author} {\bibfnamefont {V.~S.}\ \bibnamefont
  {Ilchenko}},\ }\href@noop {} {\bibfield  {journal} {\bibinfo  {journal} {J.
  Opt. Soc. Am. B}\ }\textbf {\bibinfo {volume} {16}},\ \bibinfo {pages} {147}
  (\bibinfo {year} {1999})}\BibitemShut {NoStop}%
\bibitem [{\citenamefont {Zhu}\ \emph {et~al.}(2010)\citenamefont {Zhu},
  \citenamefont {Ozdemir}, \citenamefont {Xiao}, \citenamefont {Li},
  \citenamefont {He}, \citenamefont {Chen},\ and\ \citenamefont
  {Yang}}]{zhu2010chip}%
  \BibitemOpen
  \bibfield  {author} {\bibinfo {author} {\bibfnamefont {J.}~\bibnamefont
  {Zhu}}, \bibinfo {author} {\bibfnamefont {S.~K.}\ \bibnamefont {Ozdemir}},
  \bibinfo {author} {\bibfnamefont {Y.-F.}\ \bibnamefont {Xiao}}, \bibinfo
  {author} {\bibfnamefont {L.}~\bibnamefont {Li}}, \bibinfo {author}
  {\bibfnamefont {L.}~\bibnamefont {He}}, \bibinfo {author} {\bibfnamefont
  {D.-R.}\ \bibnamefont {Chen}}, \ and\ \bibinfo {author} {\bibfnamefont
  {L.}~\bibnamefont {Yang}},\ }\href@noop {} {\bibfield  {journal} {\bibinfo
  {journal} {Nat. Photonics}\ }\textbf {\bibinfo {volume} {4}},\ \bibinfo
  {pages} {46} (\bibinfo {year} {2010})}\BibitemShut {NoStop}%
\bibitem [{\citenamefont {Ward}\ \emph {et~al.}(2014)\citenamefont {Ward},
  \citenamefont {Maimaiti}, \citenamefont {Le},\ and\ \citenamefont {{Nic
  Chormaic}}}]{ward_rsi_2014}%
  \BibitemOpen
  \bibfield  {author} {\bibinfo {author} {\bibfnamefont {J.~M.}\ \bibnamefont
  {Ward}}, \bibinfo {author} {\bibfnamefont {A.}~\bibnamefont {Maimaiti}},
  \bibinfo {author} {\bibfnamefont {V.~H.}\ \bibnamefont {Le}}, \ and\ \bibinfo
  {author} {\bibfnamefont {S.}~\bibnamefont {{Nic Chormaic}}},\ }\href@noop {}
  {\bibfield  {journal} {\bibinfo  {journal} {Rev. Sci. Instrum.}\ }\textbf
  {\bibinfo {volume} {85}},\ \bibinfo {pages} {111501} (\bibinfo {year}
  {2014})}\BibitemShut {NoStop}%
\bibitem [{\citenamefont {Shu}\ \emph {et~al.}(2018)\citenamefont {Shu},
  \citenamefont {Jiang}, \citenamefont {Zhao},\ and\ \citenamefont
  {Yang}}]{shu2018scatterer}%
  \BibitemOpen
  \bibfield  {author} {\bibinfo {author} {\bibfnamefont {F.}~\bibnamefont
  {Shu}}, \bibinfo {author} {\bibfnamefont {X.}~\bibnamefont {Jiang}}, \bibinfo
  {author} {\bibfnamefont {G.}~\bibnamefont {Zhao}}, \ and\ \bibinfo {author}
  {\bibfnamefont {L.}~\bibnamefont {Yang}},\ }\href@noop {} {\bibfield
  {journal} {\bibinfo  {journal} {Nanophotonics}\ }\textbf {\bibinfo {volume}
  {7}},\ \bibinfo {pages} {1455} (\bibinfo {year} {2018})}\BibitemShut
  {NoStop}%
\bibitem [{\citenamefont {Ward}\ \emph {et~al.}(2018)\citenamefont {Ward},
  \citenamefont {Yang}, \citenamefont {Lei}, \citenamefont {Yu}, \citenamefont
  {Xiao},\ and\ \citenamefont {Nic~Chormaic}}]{Ward:18}%
  \BibitemOpen
  \bibfield  {author} {\bibinfo {author} {\bibfnamefont {J.~M.}\ \bibnamefont
  {Ward}}, \bibinfo {author} {\bibfnamefont {Y.}~\bibnamefont {Yang}}, \bibinfo
  {author} {\bibfnamefont {F.}~\bibnamefont {Lei}}, \bibinfo {author}
  {\bibfnamefont {X.-C.}\ \bibnamefont {Yu}}, \bibinfo {author} {\bibfnamefont
  {Y.-F.}\ \bibnamefont {Xiao}}, \ and\ \bibinfo {author} {\bibfnamefont
  {S.}~\bibnamefont {Nic~Chormaic}},\ }\href@noop {} {\bibfield  {journal}
  {\bibinfo  {journal} {Optica}\ }\textbf {\bibinfo {volume} {5}},\ \bibinfo
  {pages} {674} (\bibinfo {year} {2018})}\BibitemShut {NoStop}%
\end{thebibliography}%
% \bibliographyfullrefs{reff}
\end{document}